\documentclass[pra,amsmath,twocolumn,floatfix,showpacs]{revtex4}
\usepackage{times}
\usepackage{graphicx}
\usepackage{amsmath}
\usepackage{mathptmx}

\newcommand{\parti}[2]{\frac{\partial #1}{\partial #2}}
\newcommand{\partit}[2]{\frac{\partial^2 #1}{\partial #2^2}}

\newcommand{\intall}{\int_{-\infty}^{\infty}}
\newcommand{\bra}[1]{\langle#1|}
\newcommand{\ket}[1]{|#1\rangle}

\begin{document}

\title{Propagation of temporal entanglement}
\author{Mankei Tsang}
\email{mankei@sunoptics.caltech.edu}
\author{Demetri Psaltis}
\date{\today}
\affiliation{
Department of Electrical Engineering, 
California Institute of Technology, Pasadena, CA 91125}
\begin{abstract}
The equations that govern the temporal evolution of two photons in the
Schr{\"o}dinger picture are derived, taking into account the effects of
loss, group-velocity dispersion, temporal phase modulation, linear
coupling among different optical modes, and four-wave mixing. Inspired
by the formalism, we propose the concept of quantum temporal imaging,
which uses dispersive elements and temporal phase modulators to
manipulate the temporal correlation of two entangled photons. We also
present the exact solution of a two-photon vector soliton, in order to
demonstrate the ease of use and intuitiveness of the proposed
formulation.
\end{abstract}
\maketitle

\section{\label{intro}Introduction}
In quantum optics, the Heisenberg picture, where optical fields are
treated as conjugate positions and momenta of quantized harmonic
oscillators, is often preferred, as it is easy to substitute the
optical fields in classical electromagnetic problems with
non-commutative operators and obtain the Heisenberg equations of
motion. Once the operator equations are solved, one can then obtain
various quantum properties of the optical fields via non-commutative
algebra. However, the Heisenberg picture is not without
shortcomings. It can be hard to analytically or numerically solve the
complex or nonlinear operator equations without approximations. It is
also difficult to grasp any intuition about how the quantum
correlations among the photons evolve until the Heisenberg equations
are solved.  These difficulties have led to a growing appreciation of
the Schr{\"o}dinger picture, where the photons are treated as an ensemble
of bosons, and the evolution of the many-photon probability amplitude
is studied. This arguably more intuitive approach has led to great
success in the quantum theory of solitons \cite{lai}, where instead of
solving the formidable nonlinear operator equations, one can obtain
analytic solutions from the \emph{linear} boson equations in the
Schr{\"o}dinger picture. The many-boson interpretation has been applied to
the study of entangled photons as well, where the two-photon
probability amplitude is shown to obey the Wolf equations by Saleh,
Teich, and Sergienko (STS) \cite{saleh}. Instead of treating the
entanglement properties of the photons and the optical propagation as
two separate problems, with the STS equations, one can now use a
single quantity, namely the two-photon amplitude, to keep track of the
spatiotemporal entanglement evolution in free space. This is analogous
to the Wolf equations, which reformulate the laws of optics in terms
of coherence propagation \cite{wolf}.

In this paper, we utilize the STS treatment of two photons to study
various temporal effects, in the hope that the Schr{\"o}dinger picture
would offer a more accessible interpretation of temporal entanglement
propagation for analytic or numerical studies of two-photon
systems. Loss, group-velocity dispersion, temporal phase modulation,
via an electro-optic modulator for example, linear mode coupling, via
a beam splitter or a fiber coupler for example, and four-wave mixing,
in a coherently prepared atomic gas \cite{lukin} for example, are all
included in our proposed formalism, thus extending the STS model for
use in many more topics in quantum optics, such as nonlocal dispersion
cancellation \cite{franson,steinberg}, fourth-order interferometry
\cite{hom}, and two-photon nonlinear optics \cite{lukin,chiao}. The
analysis of a two-photon vector soliton, consisting of two photons in
orthogonal polarizations under the cross-phase modulation effect, is
presented in the final section, in order to demonstrate the ease of
use and intuitiveness of the Schr{\"o}dinger picture.

Inspired by the formalism set forth, we propose the concept of quantum
temporal imaging, which uses dispersive elements and temporal phase
modulators to manipulate the temporal entanglement properties of two
photons. Most significantly, we show that it is possible to convert
positive time correlation to negative time correlation, or vice versa,
using a temporal imaging system. This conversion technique should be
immensely useful for applications that require negative time
correlation, such as quantum-enhanced clock synchronization
\cite{giovannetti_nature}.  Although there have been theoretical
\cite{giovannetti,walton,torresOL,tsangPRA} and experimental
\cite{kuzucu} proposals of generating negative time correlation
directly, they have various shortcomings compared with the
conventional tried-and-true schemes that generate positive time
correlation.  Our proposed technique should therefore allow more
flexibility in choosing two-photon sources for quantum
optics applications.

The paper is structured as follows: Sec.~\ref{formalism} derives the
equations that describe the evolution of the two-photon amplitude in
two separate modes, Sec.~\ref{qti} introduces the principles of
quantum temporal imaging, Sec.~\ref{coupling} includes linear mode coupling
in the formalism, Sec.~\ref{Nmodes} generalizes the formalism to two
photons in more than two modes, Sec.~\ref{fwm} includes the effect
of four-wave mixing, and Sec.~\ref{vsoliton} presents the exact
solution of a two-photon vector soliton.

\section{\label{formalism}Two Photons in Two Separate Modes}
Let us first consider two photons in two optical modes, such as two
polarizations, two propagation directions or two waveguide modes.
The corresponding two-photon wavefunction is
\begin{align}
\ket{\Psi} &=  C_{12}\ket{1,1}
+ C_{11}\ket{2,0} + C_{22}\ket{0,2},
\end{align}
where the constants $C_{jk}$'s are the overall amplitudes of the
quantum states, $\ket{1,1}$ is the quantum state in which one photon
is in each mode, $\ket{2,0}$ is the state in which both photons are in
mode 1, and $\ket{0,2}$ is the state which both photons are in mode 2.
The positive-frequency forward-propagating component of the electric field
in each mode is given by \cite{huttner,matloob}
\begin{align}
\hat{E}^{(+)}_j(z,t) &= i\int_0^\infty d\omega\mbox{ }
\Big(\frac{\hbar\omega\eta_j(\omega)}{4\pi\epsilon_0 c[n_j(\omega)]^2 S}\Big)^{\frac{1}{2}}
\times\nonumber\\&\quad 
\hat{a}_j(z,\omega) \exp(-i\omega t),
\end{align}
where $n_j$ is the complex, frequency-dependent refractive
index in mode $j$, $\eta_j$ is the real part of $n_j$, $S$
is an area  of quantization in the $x$-$y$ plane, and $\hat{a}_j$
is the photon annihilation operator, related to the corresponding
creation operator via the equal-space commutator \cite{huttner,matloob},
\begin{align}
[\hat{a}_j(z,\omega),\hat{a}_j^\dagger(z,\omega')] &= \delta(\omega-\omega'),
\quad j = 1,2.
\end{align}
In the Heisenberg picture, the creation and annihilation operators evolve
according to the following equations \cite{huttner,matloob},
\begin{align}
\parti{\hat{a}_1(z,\omega)}{z} &=
i\frac{\omega n_1(\omega)}{c}\hat{a}_1(z,\omega) +
\nonumber\\&\quad
i\Big(\frac{2\omega \mu_1(\omega)}{c}\Big)^{\frac{1}{2}}
\hat{f}_1(z,\omega),
\\
\parti{\hat{a}_2(z',\omega')}{z'} &=
i\frac{\omega' n_2(\omega')}{c}\hat{a}_2(z',\omega') +
\nonumber\\&\quad
i\Big(\frac{2\omega'\mu_2(\omega')}{c}\Big)^{\frac{1}{2}}
\hat{f}_2(z',\omega'),
\end{align}
where $\mu_j$ is the imaginary part of
$n_j$, and $\hat{f}_j$ is the Langevin noise operator, satisfying
the commutation relation,
\begin{align}
[\hat{f}_j(z,\omega),\hat{f}_j^\dagger(z',\omega')]
&= \delta(z-z')\delta(\omega-\omega').
\end{align}
To proceed, we replace $\omega n_j(\omega)/c$ by the following
phenomenological approximation \cite{agrawal},
\begin{align}
\frac{\omega n_j(\omega)}{c}&\approx
i\frac{\alpha_j}{2}+
\sum_{n=0}^2\frac{\beta_{nj}}{n!}(\omega-\omega_0)^n
+\frac{\omega_0}{c}\Delta n_j,
\label{kapprox}
\end{align}
where $\alpha_j = 2Im[k_j(\omega_0)]$ is the loss coefficient,
$\beta_{nj} = \partial^n Re[k_j(\omega)]/\partial\omega^n
|_{\omega=\omega_0}$ is the $n$th-order dispersion coefficient, and
$\Delta n_j$ encompasses any other refractive index perturbation.
Defining the slowly-varying envelope operators as
\begin{align}
\hat{A}_j(z,t) &= \exp(-i\beta_{0j}z+i\omega_0 t)
\int_0^\infty \frac{d\omega}{\sqrt{2\pi}}\mbox{ }
\hat{a}_j(z,\omega)\exp(-i\omega t),
\end{align}
where $\omega_0$ is the carrier frequency of the two modes,
one can obtain two evolution equations for the envelope operators,
\begin{align}
\parti{}{z}\hat{A}_1(z,t) &= iK_1\Big(t,i\parti{}{t}\Big)
\hat{A}_1(z,t)+\hat{F}_1,
\label{heisenberg1}\\
\parti{}{z'}\hat{A}_2(z',t') &= iK_2\Big(t',i\parti{}{t'}\Big)
\hat{A}_2(z',t')+\hat{F}_2,
\label{heisenberg2}\\
K_j\Big(t,i\parti{}{t}\Big) &=\Big[\frac{i\alpha_j}{2}+i\beta_{1j}\parti{}{t}
-\frac{\beta_{2j}}{2}\partit{}{t}+\frac{\omega_0}{c}\Delta n_j(t)\Big],
\end{align}
where $\hat{F}_j$ is defined as,
\begin{align}
\hat{F}_j(z,t) &= 
\exp(-i\beta_{0j}z+i\omega_0 t)
\times\nonumber\\&\quad
\int_0^\infty\frac{d\omega}{\sqrt{2\pi}}\mbox{ }
i\Big(\frac{2\omega\mu_j(\omega)}{c}\Big)^{\frac{1}{2}}
\hat{f}_j(z,\omega)
\exp(-i\omega t),
\end{align}
and $K_j$ is the complex wavenumber for the slowly-varying envelope.
$\Delta n_j$ can explicity depend on time, if the perturbation is
much slower than the optical-frequency oscillation so that
an adiabatic approximation can be made, such as in an electro-optic
modulator.

We now define the two-photon probability amplitudes as
\begin{align}
\psi_{12}(z,t,z',t')
&=\bra{0}\hat{A}_1(z,t)\hat{A}_2(z',t')\ket{\Psi},
\label{psi12def}\\
\psi_{11}(z,t,z',t')
&=\frac{1}{\sqrt{2}}\bra{0}\hat{A}_1(z,t)\hat{A}_1(z',t')\ket{\Psi},
\\
\psi_{22}(z,t,z',t')
&=\frac{1}{\sqrt{2}}\bra{0}\hat{A}_2(z,t)\hat{A}_2(z',t')\ket{\Psi}.
\end{align}
The physical significance of each amplitude
$\psi_{jk}$ is that its magnitude squared gives the probability
density, $P_{jk}$, of coincidentally measuring one photon in mode $j$
at $(z,t)$ and another photon in mode $k$ at $(z',t')$,
\begin{align}
P_{jk}(z,t,z',t') &=|\psi_{jk}(z,t,z',t')|^2.
\end{align}
Temporal entanglement is defined as the irreducibility of $|\psi_{12}|^2$
into a product of one-photon amplitudes in the form of
$a(t)b(t')$. This means that the probability of detecting a photon in
mode $1$ at time $t$ is correlated to the probability of detecting a
photon in mode $2$ at $t'$. The most popular ways of generating
entangled photons are spontaneous parametric down conversion
\cite{klyshko} and four-wave mixing \cite{fiorentino}, where the wave
mixing geometry and the spatiotemporal profile of the pump beam
determine the initial $\psi_{12}$.

To obtain the evolution equations for the two-photon amplitude
$\psi_{12}(z,t,z',t')$ in the Schr{\"o}dinger picture, we employ the
same trick as in Ref.~\cite{saleh}. First we multiply
Eq.~(\ref{heisenberg1}) with $\hat{A}_2(z',t')$ and
Eq.~(\ref{heisenberg2}) with $\hat{A}_1(z,t)$ to produce two
equations,
\begin{align}
\parti{}{z}\hat{A}_1(z,t)\hat{A}_2(z',t')
&=iK_1\Big(t,i\parti{}{t}\Big)\hat{A}_1(z,t)\hat{A}_2(z',t')+
\hat{F}_1\hat{A}_2,
\\
\parti{}{z'}\hat{A}_1(z,t)\hat{A}_2(z',t')
&=iK_2\Big(t',i\parti{}{t'}\Big)\hat{A}_1(z,t)\hat{A}_2(z',t')+
\hat{F}_2\hat{A}_1.
\end{align}
Using the definition of $\psi_{12}$ in Eq.~(\ref{psi12def}) and
assuming that the thermal reservoirs are in the vacuum state so that
the Langevin operators evaluate to zero when applied to the wavefunction
\cite{jeffers}, a pair of equations in terms of  $\psi_{12}$ are derived,
\begin{align}
\parti{}{z}\psi_{12}(z,t,z',t')
&=iK_1\Big(t,i\parti{}{t}\Big)
\psi_{12}(z,t,z',t'),
\label{temporal_STS1}\\
\parti{}{z'}\psi_{12}(z,t,z',t')
&=iK_2\Big(t',i\parti{}{t'}\Big)
\psi_{12}(z,t,z',t').
\label{temporal_STS2}
\end{align}
Equations (\ref{temporal_STS1}) and (\ref{temporal_STS2}) are the
temporal version of the STS equations \cite{saleh}, including the effects
of loss, dispersion and phase modulation. They can also be written
in the frequency domain as
\begin{align}
\phi_{12}(z,\Omega,z',\Omega') &=
\intall dt\intall dt'\mbox{ }\psi_{12}(z,t,z',t')\times\nonumber\\
&\quad \exp(i\Omega t+i\Omega' t'),
\\
\parti{}{z}\phi_{12}(z,\Omega,z',\Omega')
&=iK_1\Big(\frac{1}{i}\parti{}{\Omega},\Omega\Big)
\phi_{12}(z,\Omega,z',\Omega'),
\label{spectral_STS1}\\
\parti{}{z'}\phi_{12}(z,\Omega,z',\Omega')
&=iK_2\Big(\frac{1}{i}\parti{}{\Omega'},\Omega'\Big)
\phi_{12}(z,\Omega,z',\Omega').
\label{spectral_STS2}
\end{align}
For entangled photons, because $\psi_{12}$ or $\phi_{12}$ cannot be
separated into a product of one-photon amplitudes, distortions
experienced in one arm can coherently add to the distortions
experienced in the other arm, leading to various nonlocal quantum
effects.

For example, consider group-velocity dispersion only, the output
$\phi_{12}$ is given by
\begin{align}
&\quad \phi_{12}(z,\Omega,z',\Omega')\nonumber\\
&=\exp\Big(i\beta_{11}\Omega z+i\beta_{12}\Omega'z'+
\frac{i\beta_{21}}{2}\Omega^2 z+
\frac{i\beta_{22}}{2}\Omega'^2 z'\Big)\times\nonumber\\
&\quad \phi_{12}(0,\Omega,0,\Omega').
\end{align}
If the photons are initially entangled with negative frequency
correlation, $\phi_{12}(0,\Omega,0,\Omega')$ can be approximated
by $\phi(\Omega)\delta(\Omega+\Omega')$. Ignoring the unimportant
linear spectral phase, the output is
\begin{align}
\phi_{12}(z,\Omega,z',\Omega') &=
\exp\Big[\frac{i\Omega^2}{2} (\beta_{21}z+
\beta_{22}z')\Big]\phi(\Omega)\delta(\Omega+\Omega').
\end{align}
Hence if $\beta_{21}z =-\beta_{22}z'$, the dispersion effects in both arms
can nonlocally cancel each other, as originally discovered by Franson
\cite{franson}.

\section{\label{qti}Quantum Temporal Imaging}
In the Schr{\"o}dinger picture, the two-photon amplitude evolves under
temporal effects. Since the entanglement properties of the photons are
contained in the two-photon amplitude, the Schr{\"o}dinger picture allows
one to use the temporal effects to engineer the entanglement.

First, consider the evolution of the two-photon amplitude when
one of the modes, say mode 1,  is subject to group-velocity dispersion,
\begin{align}
\parti{\psi_{12}}{z} &= -\beta_{11}\parti{\psi_{12}}{t}
-\frac{i\beta_{21}}{2}\partit{\psi_{12}}{t},\\
\psi_{12}(L,t,z',t') &= \intall d\tau \mbox{ }
b_1(t-\tau)\psi_{12}(0,\tau,z',t'),\\
b_1(t-\tau) &=\Big(\frac{i}{2\pi \beta_{21} L}\Big)^{\frac{1}{2}} 
\exp\Big[\frac{-i(t-\beta_{11}L-\tau)^2}{2\beta_{21} L}\Big].
\end{align}
Group-velocity dispersion is well known to be analogous to Fresnel
diffraction.

Next, consider a quadratic temporal modulation of refractive index
imposed on mode 1 by a short or traveling-wave electro-optic modulator,
\begin{align}
\parti{\psi_{12}}{z} &= \frac{ik_0\Delta n_2(t-t_0)^2}{2}  \psi_{12},\\
\psi_{12}(l,t,z',t') &= q(t)\psi_{12}(0,t,z',t'),\\
q(t) &=\exp\Big[\frac{ik_0\Delta n_2 l}{2}(t-t_0)^2 \Big].
\end{align}
Quadratic temporal phase modulation is analogous to a lens.  $\Delta
n_2$ is assumed to be a constant, and $t_0$ is the time delay of the
modulation. Kerr effect by a co-propagating classical pulse would also
suffice.

Two dispersive elements and a quadratic phase modulator in-between
form a temporal imaging system, which has been well studied in the
classical domain \cite{kolner}. Suppose that the photon in mode 1
propagates through the first dispersive element, with an effective
dispersion coefficient $\beta_{21}$ and effective length $L$, then passes
through a time lens with refractive index modulation $\Delta n_2
(t-t_0)^2/2$, and finally propagates through the second dispersive
element, with an effective dispersion coefficient $\beta_{21}'$ and
effective length $L'$. The output two-photon amplitude can be expressed 
in terms of the input as
\begin{align}
\psi_{12}(z,t,z',t') &= \intall d\tau\intall d\tau'\mbox{ }
b_1'(t-\tau')q(\tau')b_1(\tau'-\tau)\times\nonumber\\
&\quad \psi_{12}(0,\tau,z',t'),
\\
b_1'(t-\tau') &=\Big(\frac{i}{2\pi \beta_{21}' L'}\Big)^{\frac{1}{2}} 
\exp\Big[\frac{-i(t-\beta_{11}'L'-\tau')^2}{2\beta_{21}' L'}\Big].
\end{align}
When the ``lens law'' for the time domain is satisfied,
\begin{align}
\frac{1}{\beta_{21} L} + \frac{1}{\beta_{21}' L'} &= k_0\Delta n_2 l,
\label{lensformula}
\end{align}
the impulse response of the system becomes
\begin{align}
h(t,\tau) &= \intall d\tau'\mbox{ }
b_1'(t-\tau')q(\tau')b_1(\tau'-\tau)\\
&=\frac{i}{2\pi\sqrt{\beta_{21}L\beta_{21}'L'}}
\exp\Big[\frac{-i(t-\beta_{11}'L')^2}{2\beta_{21}'L'}\Big]\times\nonumber\\
&\quad \exp\Big[\frac{-i(\tau+\beta_{11}L)^2}{2\beta_{21}L}\Big]
\intall d\tau' \mbox{ } P(\frac{\tau'}{T_a})\times\nonumber\\
&\quad \exp\Big[i\Big(
\frac{t-\beta_{11}'L'}{\beta_{21}' L'}+
\frac{\tau+\beta_{11}L}{\beta_{21} L}-
k_0\Delta n_2 l t_0\Big)\tau'\Big],\label{impulse}
\end{align}
where $P(\tau'/T_a)$ is the normalized temporal aperture function of the
time lens that can be used to describe any deviation of the actual
temporal phase modulation from the ideal quadratic profile, such as
truncation or higher-order phase modulation, and $T_a$
is the aperture width. If
\begin{align}
T_a >> \frac{\beta_2 L}{T_0},
\end{align}
where $T_0$ is the smallest feature size of $\psi_{12}$ along the $t$
axis, the integral in Eq.~(\ref{impulse}) can be approximated by a
delta function. We then arrive at the input-output relation for the
two-photon amplitude,
\begin{align}
\psi_{12}(z,t,z',t') &= \frac{1}{\sqrt{M}}\psi_{12}(0,\frac{t-t_d}{M},z',t'),
\label{image}\\
t_d &= \beta_{11}'L'+M\beta_{11}L+(1-M) t_0,
\\
M &= -\frac{\beta_{21}'L'}{\beta_{21} L},
\end{align}
where an unimportant quadratic phase factor is omitted, $t_d$ is the
time delay of the system, and $M$ is the magnification, which can be
positive or negative depending on the signs of $\beta_2$ and
$\beta_2'$.

\begin{figure}[htbp]
\centerline{\includegraphics[width=0.46\textwidth]{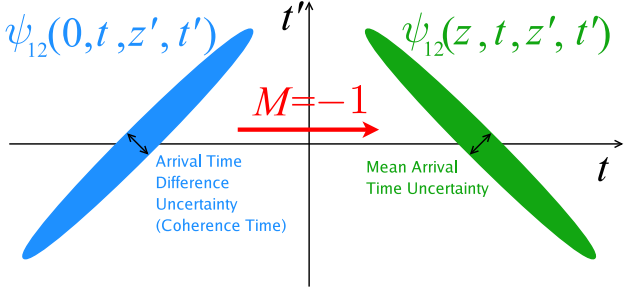}}
\caption{(color online).
Two-dimensional sketches of the two-photon probability amplitude
before and after one of the photons is time-reversed. Uncertainty in
arrival time difference is transformed to uncertainty in mean arrival
time.}
\label{time_reversal}
\end{figure}

\begin{figure}[htbp]
\centerline{\includegraphics[width=0.46\textwidth]{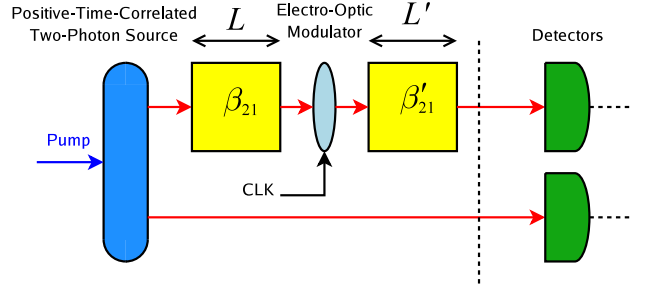}}
\caption{(color online).
A quantum temporal imaging system for quantum-enhanced
clock synchronization.}
\label{imaging}
\end{figure}

The most interesting case is when $M = -1$, and one of the photons is
time-reversed. If the two photons are initially entangled with
positive time correlation, $\psi_{12}$ can be written as
\begin{align}
\psi_{12}(0,t,z',t') &= a(t)b(t-t'),
\end{align}
where $b$ is assumed to be much sharper than $a$. After photon 1 has
passed through the temporal imaging system with $M = -1$,
\begin{align}
\psi_{12}(z,t,z',t') &= a(t_d-t)b(t_d-t-t').
\end{align}
The photons hence become anti-correlated in time. See
Fig.~\ref{time_reversal} for an illustration of this process. Since
most conventional two-photon sources generate positive time
correlation, but negative time correlation is desirable for many
applications, one can use the temporal imaging system to convert the
former to the latter. In particular, using the aforementioned
technique for the specific application of clock synchronization, the
sub-classical uncertainty of arrival time difference, $(t-t')/2$, can
be converted to a sub-classical uncertainty of mean arrival time,
$(t+t')/2$, leading to a quantum enhancement of clock synchronization
accuracy by a factor of $\sqrt{2}$ over the classical limit. In
practice, the clock can be synchronized with the electro-optic
modulator, so that the mean arrival time is controlled by $t_0$ and
thus the clock. The proposed setup is drawn in Fig.~\ref{imaging}.

The fidelity of time reversal is limited by parasitic effects, such as
higher-order dispersion and phase modulation, and the temporal
aperture $T_a$, which adds a factor $\sim \beta_{21}'L'/T_a$ to the width
of $\psi$ along the $t$ axis and increases the overall uncertainty of the
mean arrival time. The ultimate limit, apart from instrumental ones, is
set by the failure of the slowly-varying envelope approximation, which
only concerns ultrashort pulses with few optical cycles.

Besides the above application, one can also convert negative time
correlation, which can be generated by ultrashort pulses for improved
efficiency \cite{walton,tsangPRA,giovannettiPRA}, to positive time
correlation. As evident from Eq.~(\ref{image}), any desired
correlation can actually be imposed on already entangled photons, by
multiplying the original correlation with a factor of $1/M$.

As group-velocity dispersion and temporal phase modulation play
analogous roles in the time domain to diffraction and lenses, one can
use Fourier optics \cite{goodman}, temporal imaging \cite{kolner}, and
quantum imaging \cite{abouraddy} techniques to design more complex
quantum temporal imaging systems.

\section{\label{coupling}Two Photons in Two Linearly-Coupled Modes}
Suppose that the two modes are now coupled to each other, via, for
example, a beam splitter or a fiber coupler. Equations
(\ref{heisenberg1}) and (\ref{heisenberg2}) become coupled-mode
equations,
\begin{align}
\Big(\parti{}{z}-iK_1\Big)
\hat{A}_1&= 
i\kappa(z) \hat{A}_2 + \hat{F}_1,
\label{heisenberg1c}\\
\Big(\parti{}{z'}-iK_2'\Big)
\hat{A}_2'&= 
i\kappa^*(z') \hat{A}_1' + \hat{F}_2',
\label{heisenberg2c}
\end{align}
where $\kappa$ is the coupling coefficient, and for simplicity the
coupling is assumed to be co-directional.  The primes denote
the evaluations of the functions at $(z',t')$.  Any phase mismatch can be
incorporated into $\kappa$ as a $z$-dependent phase.

Procedures similar to those in Sec.~\ref{formalism} produce
four coupled equations for $\psi_{11}$, $\psi_{22}$, and $\psi_{12}$,
\begin{align}
\Big(\parti{}{z}-iK_1\Big)
\sqrt{2}\psi_{11}(z,t,z',t')
&= i\kappa \psi_{12}(z',t',z,t),\label{psi11}\\
\Big(\parti{}{z'}-iK_2'\Big)
\sqrt{2}\psi_{22}(z,t,z',t')
&= i{\kappa^*}' \psi_{12}(z',t',z,t),\label{psi22}\\
\Big(\parti{}{z}-iK_1\Big)
\psi_{12}(z,t,z',t')
&= i\kappa \sqrt{2}\psi_{22}(z,t,z',t'),\label{psi121}\\
\Big(\parti{}{z'}-iK_2'\Big)
\psi_{12}(z,t,z',t')
&= i{\kappa^*}' \sqrt{2}\psi_{11}(z,t,z',t').\label{psi122}
\end{align}
Any pair of Eqs.~(\ref{psi22}) and (\ref{psi121}) or
Eqs.~(\ref{psi11}) and (\ref{psi122}) can be combined to yield a
single equation for $\psi_{12}$,
\begin{align}
&\quad \Big(\parti{}{z}-iK_1\Big)
\Big(\parti{}{z'}-iK_2'\Big)
\psi_{12}(z,t,z',t')\nonumber\\
&=-\kappa(z)\kappa^*(z')\psi_{12}(z',t',z,t).
\label{tsang}
\end{align}
Equation (\ref{tsang}) allows one to calculate the coupled-mode
propagation of two photons in terms of $\psi_{12}$ only, given the
initial conditions of $\psi_{12}$, $\psi_{11}$, and
$\psi_{22}$. $\psi_{11}$ and $\psi_{22}$ can then be obtained from
Eqs.~(\ref{psi121}) and (\ref{psi122}) after $\psi_{12}$ is
calculated.

\begin{figure}[htbp]
\centerline{\includegraphics[width=0.46\textwidth]{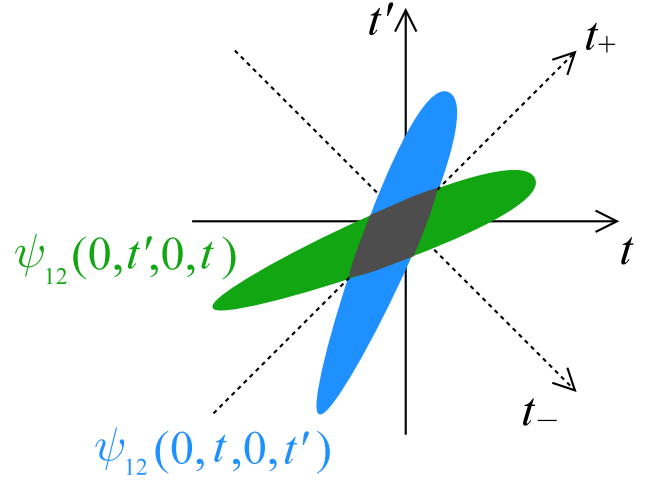}}
\caption{(color online).
The quantum destructive interference via a coupler is
determined by the overlap (dark grey area) of the two-photon amplitude
$\psi_{12}(0,t,0,t')$ with its mirror image with respect to the $t+t'$ axis,
$\psi_{12}(0,t',0,t)$.}
\label{hom}
\end{figure}

To obtain some insight into Eq.~(\ref{tsang}), consider only constant
mode coupling, so that Eq.~(\ref{tsang}) becomes
\begin{align}
\parti{}{z}\parti{}{z'}\psi_{12}(z,t,
z',t') &=
-\kappa^2\psi_{12}(z',t',
z,t).
\end{align}
The solution is
\begin{align}
\psi_{12}(z,t,z',t') 
&=\cos(\kappa z)\cos(\kappa z')\psi_{12}(0,t,0,t')-\nonumber\\
&\quad \sin(\kappa z)\sin(\kappa z')\psi_{12}(0,t',0,t)+\nonumber\\
&\quad i\sin(\kappa z)\cos(\kappa z')
\sqrt{2}\psi_{22}(0,t,0,t')+\nonumber\\
&\quad i\cos(\kappa z)\sin(\kappa z')\sqrt{2}\psi_{11}(0,t,0,t').
\end{align}
At the coupler output, $z=z'=L$,
\begin{align}
\psi_{12}(L,t,L,t')
&=T\psi_{12}(0,t,0,t')- 
R\psi_{12}(0,t',0,t)+ \nonumber\\
&\quad i\sqrt{2TR}\psi_{22}(0,t,0,t')+\nonumber\\
&\quad i\sqrt{2TR}\psi_{11}(0,t,0,t').
\end{align}
where $T = \cos^2(\kappa L)$ and $R = 1-T = \sin^2(\kappa L)$.
If we have one photon in each mode initially, only the initial
condition of $\psi_{12}$ is non-zero, and
\begin{align}
\psi_{12}(L,t,L,t')
&= T\psi_{12}(0,t,0,t')-
R\psi_{12}(0,t',0,t).
\label{destructive}
\end{align}
From Eq.~(\ref{destructive}), one can see that the output amplitude is
the destructive interference between the original amplitude and its
replica but with the two photons exchanging their positions in time.
In particular, for a 50\%-50\% coupler, $T = R = 1/2$, complete
destructive interference is produced if the two input photons are
temporally indistinguishable. See Fig.~\ref{hom} for a graphical
illustration of the destruction interference.  The introduction of
variable distinguishability to photons, in order to produce varying
degrees of destructive interference of $\psi_{12}$ via a beam splitter
and to measure the two-photon coherence time, is the basic principle
of the Hong-Ou-Mandel interferometer \cite{hom}.

\section{\label{Nmodes} Two Photons in Many Modes}
If the two photons are optically coupled to more than two modes, such
as four modes for two polarizations in each of the two propagation
directions, or $N$ modes in an array of $N$ fibers coupled to each
other, one in general needs $N(N+1)/2$ two-photon amplitudes to
describe the system. The propagation of the amplitudes in many
modes is described by the following,
\begin{align}
&\quad
\Big(\parti{}{z}-iK_j\Big)
\sqrt{1+\delta_{jk}}\psi_{jk}(z,t,z',t')
\nonumber\\
&=i\sum_{l\neq j}\kappa_{jl}\sqrt{1+\delta_{lk}}
\psi_{lk}(z,t,z',t'),
\label{many}
\end{align}
where
\begin{align}
\psi_{jk}(z,t,z',t') &=
\psi_{kj}(z',t',z,t), \quad \kappa_{jk} =
\kappa_{kj}^*.
\end{align}
Further simplications can also be made if any of the coupling terms is zero.

For example, let there be four modes; mode 1 corresponds to arm 1
with $x$ polarization, mode 2 corresponds to arm 2 with $x$ polarization,
mode 3 corresponds to arm 1 with $y$ polarization, and mode
4 corresponds to arm 2 with $y$ polarization. If only the same
polarizations are coupled, the two-photon equations are
\begin{align}
&\left(\begin{array}{cccc}
\parti{}{z} & -i\kappa_x & 0 & 0\\
-i\kappa_x & \parti{}{z} & 0 & 0\\
0 & 0 & \parti{}{z} & -i\kappa_y\\
0 & 0 & -i\kappa_y & \parti{}{z}
\end{array}\right)\times\nonumber\\
&\left(\begin{array}{cccc}
\sqrt{2}\psi_{11} & \psi_{12} & \psi_{13} & \psi_{14}\\
\psi_{21} & \sqrt{2}\psi_{22} & \psi_{23} & \psi_{24}\\
\psi_{31} & \psi_{32} & \sqrt{2}\psi_{33} & \psi_{34}\\
\psi_{41} & \psi_{42} & \psi_{43} & \sqrt{2}\psi_{44}
\end{array}\right) = 0.
\end{align}
The following solution for the orthogonally polarized amplitudes can
be obtained,
\begin{align}
&\quad \left(\begin{array}{c}
\psi_{13}(L,t,L,t') \\
\psi_{24}(L,t,L,t') \\
\psi_{14}(L,t,L,t') \\
\psi_{23}(L,t,L,t') 
\end{array}\right)\nonumber\\&= 
\left(\begin{array}{cccc}
\sqrt{T_xT_y} & -\sqrt{R_x R_y} & i\sqrt{T_xR_y} & i\sqrt{T_yR_x}\\
-\sqrt{R_x R_y} & \sqrt{T_xT_y} & i\sqrt{T_yR_x} & i\sqrt{T_xR_y}\\
i\sqrt{T_x R_y} & i\sqrt{T_yR_x} & \sqrt{T_xT_y} & -\sqrt{R_xR_y}\\
i\sqrt{T_y R_x} & i\sqrt{T_xR_y} & -\sqrt{R_xR_y} & \sqrt{T_xT_y}
\end{array}\right)\times\nonumber\\
&\quad \left(\begin{array}{c}
\psi_{13}(0,t,0,t') \\
\psi_{24}(0,t,0,t') \\
\psi_{14}(0,t,0,t') \\
\psi_{23}(0,t,0,t') 
\end{array}\right),
\end{align}
where $T_{x,y} = \cos^2(\kappa_{x,y}L)$ and $R_{x,y}=1-T_{x,y}$.
In particular, if only  the initial condition of $\psi_{14}$ is non-zero,
\begin{align}
\psi_{13}(L,t,L,t')&=i\sqrt{T_xR_y}\psi_{14}(0,t,0,t'),
\\
\psi_{24}(L,t,L,t')&=i\sqrt{T_yR_x}\psi_{14}(0,t,0,t'),
\\
\psi_{14}(L,t,L,t')&=\sqrt{T_xT_y}\psi_{14}(0,t,0,t'),
\\
\psi_{23}(L,t,L,t')&=-\sqrt{R_xR_y}\psi_{14}(0,t,0,t').
\end{align}
The singlet state for orthogonally polarized photons is produced if
$T_x = T_y = 1/2$ \cite{ohm}.

\section{\label{fwm}Four-Wave Mixing}
As envisioned by Lukin \textit{et al.}, the third-order nonlinear
effects among two photons can become significant in a coherently
prepared atomic gas \cite{lukin}. The coupled-mode equations
(\ref{heisenberg1c}) and (\ref{heisenberg2c}) then become nonlinear,
\begin{align}
\Big(\parti{}{z} -
iK_1\Big)\hat{A}_1
&=i\kappa \hat{A}_2+i
\gamma\hat{A}_1^\dagger\hat{A}_1\hat{A}_1+
i\eta\hat{A}_2^\dagger\hat{A}_2\hat{A}_1+
\nonumber\\&\quad
i\chi \hat{A}_2\hat{A}_2\hat{A}_1^\dagger + \hat{F}_1,\label{fwmheisenberg1}\\
\Big(\parti{}{z'} - iK_2'\Big)\hat{A}_2'
&=i{\kappa^*}'\hat{A}_1'+
i\gamma\hat{A'}_2^{\dagger} \hat{A}_2'\hat{A}_2'+
i\eta\hat{A'}_1^{\dagger}\hat{A}_1'\hat{A}_2'+
\nonumber\\&\quad
i{\chi^*}'\hat{A}_1'\hat{A}_1'\hat{A'}_2^{\dagger} + \hat{F}_2',
\label{fwmheisenberg2}
\end{align}
where $\gamma$ is the self-phase modulation coefficient, $\eta$ is the
cross-phase modulation coefficient, and $\chi$ is the four-wave mixing
coefficient. If we define equal-space two-photon amplitudes as the
following,
\begin{align}
\psi_{jk}(z,t,t') &= \psi_{jk}(z,t,z,t'),
\end{align}
three \emph{linear} coupled-mode equations for the
two-photon amplitudes can be derived,
\begin{align}
\Big(\parti{}{z} -
iK_1-iK_1'\Big)\sqrt{2}\psi_{11} &=
i\kappa \psi_{21}+
i\gamma\delta(t-t')\sqrt{2}\psi_{11}+\nonumber\\
&\quad i\chi\delta(t-t')\sqrt{2}\psi_{22},\label{fwm11}\\
\Big(\parti{}{z} -iK_2-
iK_2'\Big)\sqrt{2}\psi_{22} &=
i{\kappa^*} \psi_{21}+
i\gamma\delta(t-t')\sqrt{2}\psi_{22}+\nonumber\\
&\quad i\chi^*\delta(t-t')\sqrt{2}\psi_{11},\label{fwm22}\\
\Big(\parti{}{z} -
iK_1-iK_2'\Big)\psi_{12} &=
i\kappa^*\sqrt{2}\psi_{11}+i\kappa \sqrt{2}\psi_{22}+\nonumber\\
&\quad i\eta\delta(t-t')\psi_{12}.\label{fwm12}
\end{align}
The advantage of the Schr{\"o}dinger picture is most evident here; whereas
in the Heisenberg picture one needs to solve nonlinear coupled-mode
operator equations such as Eqs.~(\ref{fwmheisenberg1}) and
(\ref{fwmheisenberg2}), in the Schr{\"o}dinger picture, one only needs to
solve linear equations such as Eqs.~(\ref{fwm11}) to (\ref{fwm12}),
which are similar to the configuration-space model applied to the
quantum theory of solitons \cite{lai,hagelstein}.

The delta function $\delta(t-t')$ couples the two subspaces of
$\psi_{12}(z,t,t')$, so entanglement can emerge from unentangled
photons \cite{lukin}. To see this effect, assume that we only have 
four-wave mixing, so that Eq.~(\ref{fwm12}) becomes
\begin{align}
\parti{}{z}\psi_{12}(z,t,t') &=
i\eta\delta(t-t')
\psi_{12}(z,t,t'),
\label{fwmtime1}
\end{align}
which yields
\begin{align}
\psi_{12}(L,t,t')&=\exp[i\eta L\delta(t-t')]\psi_{12}(0,t,t').
\label{zplus}
\end{align}
If the nonlinearity has a finite bandwidth $\Delta\omega$, the delta
function in time should be replaced by a finite-bandwidth function,
for example a sinc function,
\begin{align}
\psi_{12}(L,t,t') &=\exp\Big\{\frac{i\eta L}{\pi(t-t')}
\sin\Big[\frac{\Delta\omega}{2}(t-t')\Big]\Big\}\times\nonumber\\
&\quad\psi_{12}(0,t,t').
\label{entangle}
\end{align}
Eq.~(\ref{entangle}) is the exact solution of the two-photon amplitude
under the cross-phase modulation effect, while Eq.~(7) in
Ref.~\cite{lukin}, presumably derived in the Heisenberg picture, is
only correct in the first-order. As $\psi_{12}(L,t,t')$ cannot be
written as a product of one-photon amplitudes even if the two photons
are initially unentangled, entanglement is generated.  The physical
interpretation is that the two input photons act as pump photons to
the spontaneous four-wave mixing process and are annihilated to
generate two new entangled photons.

Unlike temporal imaging techniques, which can only manipulate
the two-photon amplitude along the horizontal axis $t$ or the vertical
axis $t'$, cross-phase modulation allows some manipulation of the
two-photon amplitude along the diagonal time-difference axis, $t-t'$.
Unfortunately, cross-phase modulation by itself cannot generate any temporal
correlation, as it only imposes a phase on the two-photon temporal
amplitude. In order to have more control along the $t-t'$ axis, one can
combine the effects of cross-phase modulation and dispersion, as
shown in the following section.

\section{\label{vsoliton} Two-Photon Vector Solitons}
In this section we study a toy example, namely, a soliton formed by
two photons in orthogonal polarizations exerting cross-phase
modulation on each other \cite{agrawal}. Although similar studies of
two photons in the same mode under the self-phase modulation effect
have been performed in Refs.~\cite{chiao}, cross-phase modulation
offers the distinct possibility of entangling two photons in different
modes.

Consider the case in which two polarizations have the same
group-velocity dispersion, so that $\beta_{21}=\beta_{22} =\beta_2$,
and there is one photon in each polarization. The evolution equation
for $\psi_{12}(z,t,t')$ is
\begin{align}
&\quad
\Big(\parti{}{z}+\beta_{11}\parti{}{t}+\beta_{12}\parti{}{t'}\Big)\psi_{12}
\nonumber\\
&=\Big[-\frac{i\beta_{2}}{2}\Big(\partit{}{t}
+\partit{}{t'}\Big)+i\eta\delta(t-t')\Big]\psi_{12}.
\label{schro1}
\end{align}
Defining time coordinates in a moving frame,
\begin{align}
\tau &= t-\bar{\beta}_1z, &
\tau' &= t'-\bar{\beta}_1z,\\
\bar{\beta}_1 &= \frac{\beta_{11}+\beta_{12}}{2}, &
\Delta &=\frac{\beta_{11}-\beta_{12}}{2},
\end{align}
we obtain the following equation for $\psi_{12}(z,\tau,\tau')$,
\begin{align}
&\quad\Big(\parti{}{z}+\Delta\parti{}{\tau}-\Delta\parti{}{\tau'}\Big)\psi_{12}
\nonumber\\
&=\Big[-\frac{i\beta_{2}}{2}\Big(\partit{}{\tau}
+\partit{}{\tau'}\Big)+i\eta\delta(\tau-\tau')\Big]
\psi_{12}.\label{linearschro}
\end{align}
Equation (\ref{linearschro}) is a simple linear Schr{\"o}dinger equation,
describing a two-dimensional ``wavefunction''
$\psi_{12}(z,\tau,\tau')$ in a moving frame subject to a delta
potential. To solve for $\psi_{12}$ explicity, we define new time
coordinates,
\begin{align}
\tau_+ &= \frac{\tau+\tau'}{2},
\quad
\tau_- = \frac{\tau-\tau'}{2},
\end{align}
Eq.~(\ref{linearschro}) then becomes
\begin{align}
&\quad\Big(\parti{}{z}+\Delta\parti{}{\tau_-}\Big)\psi_{12}
\nonumber\\
&=\Big[-\frac{i\beta_{2}}{4}\Big(\partit{}{\tau_+}
+\partit{}{\tau_-}\Big)+ \frac{i\eta}{2}\delta(\tau_-)\Big]
\psi_{12}.\label{newschro}
\end{align}
As evident from Eq.~(\ref{newschro}), the cross-phase modulation
effect only offers confinement of $\psi_{12}$ along the time
difference ($\tau_-$) axis, but not the mean arrival time ($\tau_+$)
axis.

\begin{figure}[htbp]
\centerline{\includegraphics[width=0.46\textwidth]{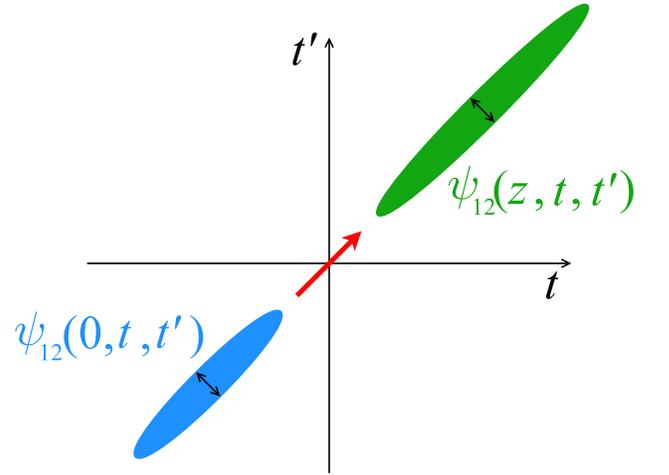}}
\caption{(color online).
Quantum dispersive spreading of mean arrival time of a
two-photon vector soliton. The cross-phase modulation effect only
preserves the two-photon coherence time, giving rise to temporal
entanglement with positive time correlation. One can also manipulate the
coherence time independently by adiabatically changing the
nonlinear coefficient along the propagation axis.}
\label{soliton}
\end{figure}

The only bound-state solution of $\psi_{12}$ is
\begin{align}
&\quad\psi_{12}(z,\tau_+,\tau_-) \nonumber\\
&=\exp\Big[-i\Big(\frac{\beta_2}{4}S^2+\frac{\Delta^2}{\beta_2}\Big)z\Big]
\times\nonumber\\
&\quad\exp\Big(-S|\tau_-|+i\frac{2\Delta}{\beta_2}\tau_-\Big)\times\nonumber\\
&\quad\intall\frac{d\Omega}{2\pi}\mbox{ }
\phi(\Omega)\exp\Big(-i\Omega\tau_++\frac{i\beta_2}{4}\Omega^2z\Big).
\label{bound}
\end{align}
The delta potential enforces $S$ to take on the following
value,
\begin{align}
S &= -\frac{\eta}{\beta_2},
\end{align} 
where $\eta$ and $\beta_2$ must have opposite signs.
The final solution of $\psi_{12}$ in the frame of $\tau$ and $\tau'$
is therefore
\begin{align}
&\quad\psi_{12}(z,\tau,\tau')\nonumber\\
&=\exp\Big[-i\Big(\frac{\eta^2/4+\Delta^2}{\beta_2}\Big)z\Big]\times\nonumber\\
&\quad\exp\Big[-\Big|\frac{\eta}{2\beta_2}\Big||\tau-\tau'|+
i\frac{\Delta}{\beta_2}(\tau-\tau')\Big]\times\nonumber\\
&\quad\intall\frac{d\Omega}{2\pi}\mbox{ }
\phi(\Omega)\exp\Big[-i\Omega(\frac{\tau+\tau'}{2})+\frac{i\beta_2}{4}
\Omega^2z\Big].
\end{align}
The two-photon coherence time of a vector soliton is fixed, but the
average arrival time is still subject to dispersive spreading and
becomes increasingly uncertain as they propagate. See
Fig.~\ref{soliton} for an illustration. Hence, a two-photon vector
soliton generates temporal entanglement with positive time correlation
as it propagates. Similar to the idea of soliton momentum squeezing
\cite{fini}, one can also adiabatically change $\eta$ or $\beta_2$ 
along the propagation axis to control independently the two-photon
coherence time.

Notice that the center frequencies of the two
photons are shifted slightly, by an amount of $\pm\Delta/\beta_2$, to
compensate for their group-velocity mismatch, so that they can
co-propagate at the average group velocity.  This is commonly known as
soliton trapping \cite{agrawal}.

If the nonlinearity has a finite bandwidth, then the potential becomes
a finite-bandwidth function like the one in Eq.~(\ref{entangle}), and
multiple bound-state solutions can be obtained via conventional
techniques of solving the linear Schr{\"o}dinger equation.

\section{Conclusion}
We have derived the general equations that govern the temporal
evolution of two-photon probability amplitudes in different coupled
optical modes. The formalism inspires the concept of quantum temporal
imaging, which can manipulate the temporal entanglement of photons via
conventional imaging techniques. The theory also offers an intuitive
interpretation of two-photon entanglement evolution, as demonstrated
by the exact solution of a two-photon vector soliton.  To conclude, we
expect the proposed formalism to be useful for many quantum signal
processing and communication applications.

\section{Acknowledgements}
This work was supported by the Engineering Research Centers Program of
the National Science Foundation under Award Number EEC-9402726 and the
Defense Advanced Research Projects Agency (DARPA).


\begin{thebibliography}{}
\bibitem{lai} Y. Lai and H. A. Haus,
\pra \textbf{40}, 844 (1989),
Y. Lai and H. A. Haus,
\pra \textbf{40}, 854 (1989).

\bibitem{saleh} B. E. A. Saleh, M. C. Teich, and A. V. Sergienko,
\prl \textbf{94}, 223601 (2005).

\bibitem{wolf} E. Wolf,
Nuovo Cimento \textbf{12}, 884 (1954).

\bibitem{lukin} M. D. Lukin and A. Imamoglu,
\prl \textbf{84}, 1419 (2000).

\bibitem{franson} J. D. Franson,
\pra \textbf{45}, 3126 (1992).

\bibitem{steinberg} A. M. Steinberg, P. G. Kwiat and R. Y. Chiao,
\pra \textbf{45}, 6659 (1992).

\bibitem{hom} C. K. Hong, Z. Y. Ou, and L. Mandel,
\prl \textbf{59}, 2044 (1987).

\bibitem{chiao} R. Y. Chiao, I. H. Deutsch, and J. C. Garrison,
\prl \textbf{67}, 1399 (1991),
I. H. Deutsch, R. Y. Chiao, and J. C. Garrison,
\prl \textbf{69} 3627 (1992).

\bibitem{giovannetti_nature} V. Giovannetti, S. Lloyd, and L. Maccone,
\nat \textbf{412}, 417 (2001).

\bibitem{giovannetti} V. Giovannetti, L. Maccone, J. H. Shapiro,
and F. N. C. Wong,
\prl \textbf{88}, 183602 (2002).

\bibitem{walton} Z. D. Walton, M. C. Booth, A. V. Sergienko,
B. E. A. Saleh, and M. C. Teich,
\pra \textbf{67}, 053810 (2003).

\bibitem{torresOL} J. P. Torres, F. Macia, S. Carrasco, and L. Torner,
\ol \textbf{30}, 314 (2005).

\bibitem{tsangPRA} M. Tsang and D. Psaltis,
\pra \textbf{71}, 043806 (2005).

\bibitem{kuzucu} O. Kuzucu, M. Fiorentino, M. A. Albota,
F. N. C. Wong, and F. X. K{\"a}rtner,
\prl {\bf 94}, 083601 (2005).

\bibitem{huttner} B. Huttner and S. M. Barnett,
\pra \textbf{46}, 4306 (1992).

\bibitem{matloob} R. Matloob, R. Loudon, S. M. Barnett, and J. Jeffers,
\pra \textbf{52}, 4823 (1995).

\bibitem{agrawal} G. P. Agrawal,
\textit{Nonlinear Fiber Optics}
(Academic Press, San Diego, 2001).


\bibitem{klyshko} M. H. Rubin, D. N. Klyshko, Y. H. Shih, and A. V. Sergienko,
\pra \textbf{50}, 5122 (1994).

\bibitem{fiorentino} M. Fiorentino, P. L. Voss, J. E. Sharping, and P. Kumar,
IEEE Photon.\ Tech.\ Lett.\ \textbf{14}, 983 (2002).

\bibitem{jeffers} J. Jeffers and S. M. Barnett,
\pra \textbf{47}, 3291 (1993).


\bibitem{kolner}
 B. H. Kolner and M. Nazarathy,
\ol \textbf{14}, 630 (1989),
B. H. Kolner,
\jqe \textbf{30}, 1951 (1994).

\bibitem{giovannettiPRA} V. Giovannetti, L. Maccone, J. H. Shapiro,
and F. N. C. Wong,
\pra \textbf{66}, 043813 (2002).

\bibitem{goodman} J. W. Goodman,
\textit{Introduction to Fourier Optics} (McGraw-Hill, Boston, 1996).

\bibitem{abouraddy} A. F. Abouraddy, B. E. A. Saleh, A. V. Sergienko,
and M. C. Teich,
\josab \textbf{19}, 1174 (2002).


\bibitem{ohm} Z. Y. Ou, C. K. Hong, and L. Mandel,
\oc \textbf{63}, 118 (1987).

\bibitem{hagelstein} P. L. Hagelstein,
\pra \textbf{54}, 2426 (1996).

\bibitem{fini} J. M. Fini and P. L. Hagelstein,
\pra \textbf{66}, 033818 (2002).

\end{thebibliography}
\end{document}